\documentclass[conference]{IEEEtran}
\usepackage[T1]{fontenc}
\usepackage[utf8]{inputenc}
\usepackage{amsmath,amssymb,amsthm}
\usepackage{microtype}
\usepackage{booktabs}
\usepackage{multirow}
\usepackage{tabularx}
\usepackage{array}

\newcolumntype{L}[1]{>{\raggedright\arraybackslash}p{#1}}
\newcolumntype{C}[1]{>{\centering\arraybackslash}p{#1}}

\usepackage{enumitem}
\usepackage{xcolor}
\usepackage{hyperref}
\hypersetup{
  colorlinks=true,
  linkcolor=blue,
  citecolor=blue,
  urlcolor=blue
}

\newtheorem{theorem}{Theorem}

\newtheorem{definition}[theorem]{Definition}
\newtheorem{example}[]{Example}
\newtheorem{remark}[theorem]{Remark}

\newcommand{\pqsafe}{\textsf{Q-Safe}}
\newcommand{\pqunsafe}{\textsf{Q-Unsafe}}
\newcommand{\pqweak}{\textsf{Q-Weakened}}
\newcommand{\classun}{\textsf{C-Unsafe}}
\newcommand{\statusset}{\mathcal{S}}
\newcommand{\layerset}{\mathcal{A}}

\newcommand{\seffconf}{\sigma_{\mathrm{eff}}^{\mathrm{conf}}}
\newcommand{\seffauth}{\sigma_{\mathrm{eff}}^{\mathrm{auth}}}
\newcommand{\schainconf}{\sigma_{\mathrm{chain}}^{\mathrm{conf}}}
\newcommand{\schainauth}{\sigma_{\mathrm{chain}}^{\mathrm{auth}}}
\newcommand{\schainmeta}{\sigma_{\mathrm{chain}}^{\mathrm{meta}}}

\title{Post-Quantum Cryptographic Analysis of Message Transformations Across the Network Stack}
\author{
\IEEEauthorblockN{Ashish Kundu, Vishal Chakraborty, Ramana Kompella}
\IEEEauthorblockA{Cisco Research\\
\{ashkundu, veec, rkompell\}@cisco.com}
}
\begin{document}
\maketitle

\begin{abstract}
When a user sends a message over a wireless network, the message does not travel as-is. It is encrypted, authenticated, encapsulated, and transformed as it descends the protocol stack from the application layer to the physical medium. Each layer may apply its own cryptographic operations using its own algorithms, and these algorithms differ in their vulnerability to quantum computers. The security of the overall communication depends not on any single layer but on the \emph{composition} of transformations across all layers.

We develop a preliminary formal framework for analyzing these cross-layer cryptographic transformations with respect to post-quantum cryptographic (PQC) readiness. We classify every per-layer cryptographic operation into one of four quantum vulnerability categories, define how per-layer PQC statuses compose across the full message transformation chain, and prove that this composition forms a bounded lattice with confidentiality composing via the join (max) operator and authentication via the meet (min). We apply the framework to five communication scenarios spanning Linux and iOS platforms, and identify several research challenges. Among our findings: WPA2-Personal provides strictly better PQC posture than both WPA3-Personal and WPA2-Enterprise; a single post-quantum layer suffices for payload confidentiality but \emph{every} layer must migrate for complete authentication; and metadata protection depends solely on the outermost layer.
\end{abstract}


\section{Introduction}
\label{sec:intro}

Consider a user on an iOS device sending an iMessage to another iOS user over a Wi-Fi network. 
The message begins as plaintext at the application layer. Before it reaches the wireless medium, three independent cryptographic systems act on it in sequence. 
First, Apple's PQ3 protocol encrypts and signs the message using a hybrid key encapsulation mechanism that combines Kyber-1024 (a post-quantum algorithm based on lattice problems) with P-256 ECDH (a classical elliptic curve scheme). 
Next, a TLS~1.3 session to Apple's relay servers encrypts the already-encrypted message using X25519 for key exchange and AES-256-GCM for symmetric encryption. 
Finally, the Wi-Fi layer encrypts the entire IP packet (which now contains the doubly-encrypted payload) using AES-128-CCMP, with keys derived through WPA3's SAE 
handshake (itself based on elliptic curve Diffie--Hellman).

The transmitted frame is thus triply encrypted: 
the innermost layer (PQ3) uses post-quantum key exchange, the middle layer (TLS) uses classical elliptic curve key exchange, and the outermost layer (WPA3) also uses classical elliptic curve key exchange. 
An adversary with a cryptographically relevant quantum computer could break the outer two layers using Shor's 
algorithm~\cite{shor1994} but would be unable to break the innermost layer. 
The message content remains confidential. However, by breaking the outer layers, the adversary 
learns that this device is communicating with Apple's iMessage servers at specific times, metadata that the inner layer's post-quantum protection cannot conceal.

This example illustrates that the post-quantum security posture of a network communication is not determined by any single protocol or layer. 
It is determined by the \emph{composition} of cryptographic transformations across the entire protocol stack, 
and different security properties (confidentiality, authentication, metadata protection) compose according to different rules.

\subsection{The Problem}
\label{sec:intro:problem}

The threat that quantum computing poses to public-key cryptography is well understood in isolation. 
Shor's algorithm~\cite{shor1994} efficiently factors integers and computes discrete logarithms, 
breaking RSA, Diffie--Hellman, and elliptic curve schemes. Grover's algorithm~\cite{grover1996} provides a quadratic speedup for brute-force search, effectively halving symmetric key lengths. 
The ``Harvest Now, Decrypt Later'' (HNDL) threat model~\cite{nistir8547} makes migration urgent: an adversary recording encrypted traffic today can decrypt it once quantum computers arrive.

NIST has responded by standardizing three post-quantum algorithms: ML-KEM (FIPS~203) for key encapsulation, ML-DSA (FIPS~204) for digital signatures, and SLH-DSA (FIPS~205) as a hash-based signature backup. Migration guidance documents (NIST IR~8547, NSA CNSA~2.0) prescribe timelines and algorithm choices.

What is missing, however, is a formal analysis of how quantum vulnerability plays out across the \emph{layered} structure of real network communications. Existing guidance reasons about protocols in isolation. 
For example, ``migrate TLS to post-quantum key exchange,'' ``adopt ML-DSA for code signing''. 
They do so without addressing how multiple layers of encryption and authentication interact. 
In practice, however, a network session involves cryptographic operations at the
link layer (Wi-Fi), possibly the network layer (VPN), the transport/session layer (TLS), and the application layer (end-to-end encryption). 
These layers use different algorithms, authenticate different entities, and protect different scopes of traffic. 
Their PQC statuses may differ, and understanding the system-level security requires understanding how those statuses compose.

\subsection{Our Approach}
\label{sec:intro:approach}

Our goals in this paper are threefolds:
\begin{enumerate}

\item to develop a formal model that captures how cryptographic operations at each 
network layer transform a message and how each operation's quantum vulnerability status can be classified;

\item to identify composition rules that determine the system-level PQC posture from per-layer statuses, 
and to prove that these rules have clean algebraic structure;

\item to apply this framework to platform configurations,
revealing patterns that inform PQC migration strategy
(including counterintuitive findings where stronger classical security correlates with weaker quantum security).
\end{enumerate}

\noindent We proceed by modeling the \emph{message transformation chain}: 
the sequence of cryptographic operations applied to a message as it descends the sender's protocol stack and, 
symmetrically, 
the inverse operations applied as it ascends the receiver's stack. Each operation is classified by its quantum vulnerability. 
We then define how per-layer statuses compose and 
prove that the resulting algebra is a bounded lattice in which confidentiality and authentication are dual operators.

We ground our analysis in the actual cryptographic algorithms negotiated by default in modern operating systems (Ubuntu 24.04 and iOS~17/18) over wireless networks. 
This produces verifiable claims about the PQC posture of real-world configurations.

\subsection{Contributions}
\label{sec:intro:contributions}

This paper makes the following contributions:

\textbf{(1) Formal model.} 
We define the message transformation chain $L_2 \circ L_3 \circ \cdots \circ L_7$ as a composition of per-layer cryptographic operations, each carrying a PQC status. 
We distinguish per-session operations (handshake/key exchange) from per-message operations (record-layer encryption) and 
formalize the dependency between them. 
In particular, that the effective PQC status of a symmetric cipher depends on the PQC status of the key exchange that produced its key material.

\textbf{(2) Composition algebra.} 
We prove that PQC status composition across layers forms a bounded lattice on the totally ordered set $\statusset = \{\classun < \pqunsafe < \pqweak < \pqsafe\}$, with confidentiality composing via the join (max) operator and authentication via the meet (min). 
We prove these are lattice duals: confidentiality requires breaking \emph{all} layers (redundant protection), 
while authentication requires breaking only \emph{one} layer (independent attack surfaces). 
We show that metadata protection depends solely on the outermost layer.

\textbf{(3) Case studies.} 
We analyze five scenarios. From Linux localhost to iOS-to-iOS iMessage, documenting the exact cryptographic 
algorithms at each active layer and computing per-layer and chain-level PQC statuses. 
These analyses reveal that WPA2-Personal (no public-key crypto) is more quantum-resistant than WPA3-Personal (elliptic curve SAE) and WPA2-Enterprise (certificate-based EAP-TLS), and 
that a single post-quantum deployment (Apple's PQ3) transforms the chain from \pqunsafe{} to \pqsafe{} for confidentiality.

\textbf{(4) Research challenges.} 
We formalize six research challenges on cross-layer PQC properties, 
including migration sufficiency, 
authentication necessity, 
the classical-quantum security tension, 
and metadata exposure depth.

The rest of this paper is organized as follows. Section~\ref{sec:challenges} motivates and formally states six research challenges. 
Section~\ref{sec:model} develops the formal model of message transformation chains and PQC status classification.
Section~\ref{sec:cases} presents five case studies with detailed per-layer decomposition. 
% Section~\ref{sec:results} states and proves the main results.
% Section~\ref{sec:related} surveys related work. 
% Section~\ref{sec:conclusion} concludes with directions for future research.

\section{Research Challenges}
\label{sec:challenges}

In migrating a network to a safe post-quantum cryptography status faces a series of challenges 
that existing work does not answer. 
While NIST IR 8547~\cite{nistir8547} prescribes which algorithms to adopt, it does not specify which \emph{layers} to migrate first, 
how many layers need migration at all, or 
how to reason about the security of a partially migrated stack in which some layers are post-quantum and others are not. 
These are structural questions about how cryptographic protections at different network layers interact.

In this section we formulate six research challenges. Their answers, we hope, will provide a principled basis for cross-layer PQC posture determination and
 migration planning. We answer Questions~1--3 and~5 completely in Section~5,
 provide a partial answer to Question~4, and present empirical evidence bearing on Question~6 through the case studies in Section~4.

\subsection{RC1: PQC Posture Composition}
\label{sec:rc1}

\noindent\textbf{Question 1.} \textit{Given a message transformation chain with $n$ active cryptographic layers, each carrying a PQC status for confidentiality and authentication,
do there exist closed-form composition rules that determine the chain-level PQC status for (a)~payload confidentiality,
(b)~authentication, and
(c)~metadata protection?
If so, what algebraic structure do these rules exhibit?}

Consider again, the iMessage scenario from Section~1. 
A message passes through three cryptographic layers, namely, WPA3 at Layer~2 (\pqunsafe), TLS~1.3 at
Layers~5 and 6 (\pqunsafe), and Apple's PQ3 protocol at Layer~7 (\pqsafe). What is the system-level PQC posture?

A na\"ive approach, perhaps, might average the three per-layer statuses, or 
take the worst case across layers, or 
take the best case. 
However, as we will see,
the correct answer depends on \emph{which security property} one asks about.

For \emph{confidentiality}, 
the system is \pqsafe: the innermost encryption
(PQ3, keyed via Kyber-1024) blocks decryption even after the outer two layers are broken. 
For \emph{authentication}, the
system is \pqunsafe: each layer authenticates a different entity (WPA3 authenticates the access point; TLS authenticates Apple's relay server;
PQ3 authenticates the sending user), 
and a quantum adversary who forges any one
of these can mount a man-in-the-middle attack at
that layer's scope regardless of what the other layers do. For metadata, the system is \pqunsafe:
an external observer sees only the outermost layer's headers,
so only that layer's confidentiality status determines what metadata is exposed.

This three-way divergence indicates that the per-layer PQC statuses do not compose arbitrarily.
There may be a small number of composition operators to characterize each security property that determine system-level posture from per-layer statuses
in a predictable way.
If such operators exist and can be characterized algebraically,
they would provide a foundation for automated PQC posture determination across arbitrary network configurations.

\subsection{RC2: Confidentiality Sufficiency}
\label{sec:rc2}

\noindent\textbf{Question 2.} \textit{What is the minimum number of layers that must be migrated to post-quantum key exchange to achieve \pqsafe{}
chain confidentiality?
Does the position of the migrated layer(s) matter for payload protection? And for metadata protection?}

Continuing with the iMessage example, observe that a single post-quantum
layer, for example PQ3 at Layer~7, suffices to make the entire chain \pqsafe{} for payload confidentiality,
even though the outer two layers are \pqunsafe.
The message is encrypted inside PQ3 before it reaches TLS or WPA3.
A quantum adversary who breaks WPA3 and TLS obtains the PQ3 ciphertext but cannot decrypt it,
because key exchange in PQ3 uses Kyber-1024, which is not
vulnerable to Shor's algorithm. The payload remains confidential.

The practical implication is striking. An organization may not need to migrate
every layer to achieve quantum-safe confidentiality for its data in transit. However, this observation raises further questions.
Does the observation generalize?
Does it hold for any number of layers, any layer positions, any combination of PQC statuses?
And if one layer suffices, then \emph{which} layer is it.
Migrating the outermost layer (say, Wi-Fi) would protect both payload and metadata. Indeed, every header and every byte is encrypted before it leaves the device.
Migrating an inner layer (e.g., application-layer encryption) protects the payload but leaves all outer-layer metadata exposed.
The two choices lead to very different security outcomes even though both achieve one layer of post-quantum confidentiality.

\subsection{RC3: Authentication Necessity}
\label{sec:rc3}

\noindent\textbf{Question 3.} \textit{What is the minimum set of layers that must be migrated to post-quantum signatures to achieve \pqsafe{} chain authentication?
Is it necessarily the case that all layers with public-key authentication must be migrated?}

Authentication exhibits different properties. In the iMessage scenario, all three layers use classical public-key authentication, i.e,
ECDSA-P256 for PQ3 message signing,
ECDSA or RSA certificates for TLS server authentication,
and SAE's implicit EC-based mutual authentication for WPA3.
The chain is \pqunsafe{} for authentication despite having \pqsafe{}
confidentiality.
This observation reflects a structural asymmetry between confidentiality and authentication in layered encryption.

For confidentiality, the layers form nested encryption. An adversary must go through all layers to reach plaintext,
and one quantum-resistant layer blocks the chain.
For authentication, each layer independently verifies a different entity's
identity. A quantum adversary
who derives the private key behind any one layer's certificate or signing key can impersonate that entity and insert herself as a
man-in-the-middle at that layer's scope, within that layer's trust relationship without disturbing the other layers' authentication at all.
The WPA3 access point's quantum-safe authentication does not prevent an attacker from forging the TLS server's certificate.
The TLS server's quantum-safe authentication does not prevent an attacker from forging iMessage sender identity.

If this reasoning generalizes, then achieving \pqsafe{} chain authentication requires migrating authentication at \emph{every} layer that currently uses public-key signatures.
It would be impossible to have any single-layer migration that protects the rest,
as there is for confidentiality.
This asymmetry, if it can be proved formally, has significant
consequences for migration timelines and resource allocation.

\subsection{RC4: Migration Preference}
\label{sec:rc4}

\noindent\textbf{Question 4.} \textit{Given an existing message transformation chain, what is the optimal ordering for layer-by-layer PQC migration that minimizes cumulative quantum risk?
Does the answer depend on whether the organization prioritizes confidentiality, authentication, or metadata protection?}

Suppose an organization can migrate one layer per quarter. Which layer should it migrate first? If a single layer suffices for confidentiality (RC2),
then the first migration already achieves quantum-safe payload protection.
But the choice of \emph{which} layer to migrate first matters, because different layers protect different payloads.
The outermost layer protects metadata, namely, network-level identifiers, traffic patterns, protocol fingerprints.
The innermost layer, on the other hand, provides end-to-end protection that survives intermediary compromise. For example, even if the TLS termination point or VPN gateway is untrusted, application-layer encryption protects the payload.

Authentication adds a further constraint. Since it requires all
layers (RC3),
the \emph{order} in which layers are migrated determines the window of vulnerability.
If the outermost layer is migrated first, metadata is immediately protected but authentication remains fully vulnerable until the last layer has been migrated. Contrastingly, if the innermost layer is migrated first, the application gains end-to-end confidentiality first. However, metadata remains exposed throughout the migration period.
The optimal ordering depends on which security properties the organization prioritizes,
the sensitivity of different metadata types, and the threat model (passive HNDL collection viz. a viz. active man-in-the-middle).

\subsection{RC5: Metadata Exposure}
\label{sec:rc5}

\noindent\textbf{Question 5.} \textit{For a message transformation chain with $n$ layers,
how deep does metadata exposure extend when the outermost $k$ layers are quantum-vulnerable?
Can the exposure depth $d^*$ be expressed as a closed-form function of per-layer PQC statuses?}

Breaking the outermost encryption layer does not immediately reveal the application plaintext. It reveals the following layer's headers.
In the iMessage example, a quantum adversary who breaks WPA3 sees the inner IP packet: the device's local IP address, the destination IP (Apple's iMessage relay server), the destination port (443), and the TLS record headers.
The TLS payload (containing the PQ3 ciphertext) remains encrypted.
If the adversary then breaks TLS, they see the iMessage-specific metadata:
recipient identifiers, message sizes, timestamps, and delivery receipts.
Only the PQ3 layer's encryption blocks further exposure; the message content itself remains protected.

Metadata exposure is therefore not a binary outcome.
It is \emph{incremental}: each successive layer that an adversary decrypts reveals one additional level of protocol headers. This exposure stops at the first layer whose confidentiality is \pqsafe. The depth of exposure, i.e., the number of header levels revealed, depends on how many consecutive layers from the outermost inward are quantum-vulnerable.
We want to characterize this depth precisely and understand what information is revealed at each level.

\subsection{RC6: Classical vs. Quantum Security Tension}
\label{sec:rc6}

\noindent\textbf{Question 6.} \textit{Under what conditions does
upgrading a protocol's classical security degrade its quantum security?
Is this tension inherent to any replacement of symmetric mechanisms with public-key mechanisms, or are there protocol designs that improve classical security without introducing quantum vulnerability?}

An unexpected pattern emerges from comparing Wi-Fi security modes.
WPA2-Personal uses a pre-shared key and derives all session keys through symmetric operations. It uses PBKDF2 for the pairwise master key, HMAC-SHA1 for the message integrity check, AES key wrapping for group key delivery, and AES-128-CCMP for data encryption.
No public-key cryptography is involved in the protocol. Therefore, it is not suceptible to Shor's algorithm.

Contrastingly, WPA3-Personal, designed as WPA2's more secure successor, replaces the 4-way handshake's key derivation with the SAE (Simultaneous Authentication of Equals) protocol,
which performs elliptic curve Diffie--Hellman on the P-256 curve~\cite{rfc7664}.
SAE provides a genuine classical improvement by addressing all of known weaknesses of WPA2-PSK. It resists offline dictionary attacks and provides forward secrecy. But SAE also introduces an elliptic curve operation that is a direct target for Shor's algorithm, where none existed before.

The result is that upgrading from WPA2-Personal to WPA3-Personal \emph{improves} classical security while \emph{degrading} quantum security. The the PQC posture moves from \pqweak{} (Grover-weakened symmetric crypto) to \pqunsafe{} (Shor-broken elliptic curve crypto).
The same pattern emerges when upgrading from WPA2-Personal to WPA2-Enterprise.
% per-user certificate-based authentication via EAP-TLS~\cite{rfc5216} is classically stronger than a shared passphrase, but it introduces ECDHE key exchange and RSA/ECDSA certificate verification---all vulnerable to Shor's algorithm.

We observe two concrete instances of this tension:
\begin{table*}[h!]
    \centering
    \begin{tabular}{@{}lll@{}}
    \toprule
    \textbf{Upgrade} & \textbf{Classical effect} & \textbf{Quantum effect} \\
    \midrule
    WPA2-PSK $\to$ WPA3-SAE & Better (resists offline dict.) & Worse (\pqweak $\to$ \pqunsafe) \\
    WPA2-PSK $\to$ WPA2-Ent. & Better (per-user certs) & Worse (\pqweak $\to$ \pqunsafe) \\
    \bottomrule
    \end{tabular}
    
\caption{Contrasting classical and quantum effect.}
\label{tab:placeholder}
\end{table*}
\noindent If this tension is inherent, i.e., if replacing a symmetric mechanism with a public-key mechanism to gain classical security necessarily
introduces quantum vulnerability, then it constrains the design space for transitional protocols and complicates migration planning for
organizations that must maintain backward compatibility.

\section{Formal Model}
\label{sec:model}

In this section, we develop our formal model. We begin by defining how individual cryptographic algorithms are classified with respect to quantum vulnerability (Section~\ref{sec:model:status}), 
then model the sequence of cryptographic transformations applied to a message as it traverses the protocol stack (Section~\ref{sec:model:chain}). Finally, we decompose each layer's transformation into its constituent operations 
and define what determines a layer's \emph{effective} PQC status (Sections~\ref{sec:model:roles} to \ref{sec:model:effective}). Table~\ref{tab:notation} provides a summary of all notation for easy reference.

\subsection{PQC Status Classification}
\label{sec:model:status}

We classify every instance of a cryptographic algorithm into exactly 
one of four categories, depending on how it is affected by quantum computation.
The four elements are defined as follows:

An algorithm is \textbf{\pqsafe} if no known quantum algorithm provides a meaningful advantage in breaking it. Examples include ML-KEM-768, ML-DSA-65, AES-256, ChaCha20-Poly1305, SHA-384, SHA-512, and HMAC-SHA-256.

An algorithm is \textbf{\pqweak} if Grover's algorithm~\cite{grover1996} reduces its effective security level but the residual security remains \emph{strictly above} 64~bits. The threshold matters: 64-bit security has been demonstrated to be within reach of classical brute-force attacks using modern distributed computing resources, so any post-quantum residual at or below 64~bits does not provide an adequate security margin. Examples of \pqweak{} algorithms include SHA-256 (256-bit $\to$ 128-bit effective for preimage under Grover) and HMAC-SHA1 used with 160-bit keys (160-bit $\to$ 80-bit effective). AES-192 ($\to$ 96-bit) and AES-256 ($\to$ 128-bit) remain \pqsafe{} rather than \pqweak, because their post-Grover residual security (96 and 128~bits respectively) is comfortably above any feasibility threshold.

An algorithm is \textbf{\pqunsafe} if it is efficiently broken by Shor's algorithm~\cite{shor1994} or a related quantum algorithm. This includes RSA, Diffie--Hellman, ECDH (including X25519 and P-256), ECDSA, Ed25519, and DSA---the algorithms that underpin virtually all deployed public-key cryptography.

Finally, an algorithm is \textbf{\classun} if it is already broken or inadequate against classical (non-quantum) adversaries. Examples include DES, RC4, MD5, SHA-1 for collision resistance, and RSA-1024.

\begin{definition}[PQC Status Set]
\label{def:status}
The \emph{PQC status set} is the totally ordered set
\begin{equation}
\label{eq:status}
\statusset = \{\classun < \pqunsafe < \pqweak < \pqsafe\}
\end{equation}
equipped with the natural ordering from least secure to most secure.
\end{definition}

We classify \emph{individual cryptographic operations}, not protocols. 
A single protocol such as TLS~1.3 may involve multiple algorithms 
with different PQC statuses: X25519 (\pqunsafe) for key exchange, ECDSA (\pqunsafe) for server authentication, HKDF-SHA384 (\pqsafe) for key derivation, and AES-256-GCM (\pqsafe) for record encryption. 
Second, an algorithm's PQC status depends on \emph{how it is used}, 
not just on what algorithm it is. AES-256 is \pqsafe{} as a symmetric cipher,
 but if the key that feeds it was established via a 
 \pqunsafe{} key exchange, the \emph{effective} confidentiality is \pqunsafe---because a quantum adversary can recover the key. 
 We make this notion precise in Section~\ref{sec:model:effective}.

Table~\ref{tab:status_examples} summarizes the classification for algorithms that appear in our case studies.

\begin{table*}[t]
\centering
\caption{PQC status classification for algorithms in this study.}
\label{tab:status_examples}
\footnotesize
\begin{tabular}{@{}llll@{}}
\toprule
\textbf{Algorithm} & \textbf{Role} & \textbf{Post-QC Security} & \textbf{Status} \\
\midrule
ML-KEM-768/1024 & KEX & Full (lattice-based) & \pqsafe \\
ML-DSA-65 & AUTH & Full (lattice-based) & \pqsafe \\
AES-256-GCM & ENC & 128-bit effective & \pqsafe \\
ChaCha20-Poly1305 & ENC & 128-bit effective & \pqsafe \\
SHA-384/SHA-512 & KDF/INT & 192/256-bit effective & \pqsafe \\
HMAC-SHA-256 & INT & 128-bit effective & \pqsafe \\
\midrule
SHA-256 (preimage) & KDF & 128-bit effective & \pqweak \\
HMAC-SHA1 (160-bit key) & INT & 80-bit effective & \pqweak \\
PBKDF2-SHA1 (256-bit PMK) & KDF & 128-bit effective$^*$ & \pqweak \\
\midrule
AES-128-CCMP & ENC & 64-bit effective & \pqunsafe$^\dagger$ \\
X25519 / ECDH-P256 & KEX & Broken (Shor) & \pqunsafe \\
ECDSA-P256 / Ed25519 & AUTH & Broken (Shor) & \pqunsafe \\
RSA-2048+ & KEX/AUTH & Broken (Shor) & \pqunsafe \\
DH-2048 & KEX & Broken (Shor) & \pqunsafe \\
\midrule
DES, RC4, MD5 & Various & Already broken & \classun \\
\bottomrule
\multicolumn{4}{@{}l}{\footnotesize $^*$PBKDF2 output is 256 bits; Grover reduces search over the passphrase}\\
\multicolumn{4}{@{}l}{\footnotesize \phantom{$^*$}space, but the 256-bit PMK has adequate entropy.}\\
\multicolumn{4}{@{}l}{\footnotesize $^\dagger$Grover-reduced to 64-bit, at the threshold of classical feasibility.}\\
\multicolumn{4}{@{}l}{\footnotesize \phantom{$^\dagger$}Mechanism differs from Shor.}
\end{tabular}
\end{table*}

\subsection{Message Transformation Chain}
\label{sec:model:chain}

A message originating at the application layer undergoes a 
sequence of cryptographic transformations as it descends the 
protocol stack toward the physical medium. 
We model each transformation as a function and the full sequence as a composition.

\begin{definition}[Layer Transformation]
\label{def:layer}
Let $L_i$ denote the cryptographic transformation applied at OSI layer~$i$ during message sending (encapsulation). Layer~$i$ receives a message $M_i$ from the layer above and produces output
\begin{equation}
\label{eq:layer}
M_{i-1} = L_i(M_i)
\end{equation}
which becomes the input to layer $i-1$.
\end{definition}

The index convention reflects the direction of encapsulation. 
Sending a message proceeds from higher layers (application, $L_7$) to lower layers (data link, $L_2$), so $L_i$ maps layer-$i$ data to layer-$(i{-}1)$ data. Layers that perform no cryptographic operations act as identity functions ($L_i(M) = M$ with only non-cryptographic header prepending).

\begin{definition}[Sending Chain]
\label{def:sending}
For a message $M_{\mathrm{app}}$ originating at the application layer, the \emph{sending chain} is the composition
\begin{equation}
\label{eq:sending}
M_{\mathrm{out}} = L_2(L_3(\cdots(L_7(M_{\mathrm{app}}))))
\end{equation}
where $M_{\mathrm{out}}$ is the frame transmitted on the physical medium.
\end{definition}

\begin{definition}[Receiving Chain]
\label{def:receiving}
The \emph{receiving chain} reverses the sending process. Given received frame $M_{\mathrm{out}}$,
\begin{equation}
\label{eq:receiving}
M_{\mathrm{app}} = L_7^{-1}(\cdots(L_3^{-1}(L_2^{-1}(M_{\mathrm{out}}))))
\end{equation}
where $L_i^{-1}$ denotes the inverse transformation (decryption, verification) at layer~$i$.
\end{definition}

\smallskip\textbf{Send/Receive Symmetry.} The PQC status of $L_i$ and $L_i^{-1}$ are identical, because both the sender and receiver use the same negotiated algorithm for a given session. The sender encrypts with AES-256-GCM; the receiver decrypts with AES-256-GCM. Both operations inherit the same PQC status. We therefore need only analyze the sending direction; the receiving direction is symmetric.

\begin{example}[Running Example: iOS-to-iOS iMessage]
\label{ex:imessage-chain}
In the iMessage scenario from Section~1, the sending chain has three active cryptographic layers. At Layer~7, Apple's PQ3 protocol encrypts the message: $M_6 = L_7^{\mathrm{PQ3}}(M_{\mathrm{app}})$. At Layers~5--6, a TLS~1.3 session encrypts the PQ3 ciphertext: $M_4 = L_5^{\mathrm{TLS}}(M_6)$. At Layer~2, WPA3 encrypts the entire IP packet: $M_{\mathrm{out}} = L_2^{\mathrm{WPA3}}(M_4)$. The full sending chain is
\begin{equation}
\label{eq:imessage-chain}
M_{\mathrm{out}} = L_2^{\mathrm{WPA3}}\!\left(L_5^{\mathrm{TLS}}\!\left(L_7^{\mathrm{PQ3}}(M_{\mathrm{app}})\right)\right)
\end{equation}
producing a triply-encrypted frame in which the application message is nested inside three independent layers of encryption.
\end{example}

\subsection{Active Layers and Protection Path}
\label{sec:model:active}

In a given session, not every OSI layer performs cryptographic operations. We define the subset that does as \emph{active layer set}.

\begin{definition}[Active Layer Set]
\label{def:active}
For a given communication session, the \emph{active layer set} is the ordered sequence
\begin{equation}
\label{eq:active}
\layerset = (L_{i_1}, L_{i_2}, \ldots, L_{i_n})
\end{equation}
containing exactly those layers that perform at least one cryptographic operation (key exchange, encryption, authentication, or integrity protection), ordered from outermost to innermost: $i_1 < i_2 < \cdots < i_n$.
\end{definition}

In our running example, $\layerset = (L_2, L_{5\text{-}6}, L_7)$ with $n = 3$. 
Contrast this to a simpler scenario, say, HTTPS over WPA2 with no application-layer encryption. 
Here, we have $\layerset = (L_2, L_{5\text{-}6})$ with $n = 2$. On a localhost connection with no TLS, $\layerset$ may be empty.

\subsection{Layer Decomposition and Operation Roles}
\label{sec:model:roles}

Each layer's cryptographic transformation decomposes into individual operations, each serving a specific role.

\begin{definition}[Layer Decomposition]
\label{def:decomp}
The transformation at layer $L_i$ decomposes into $k_i$ individual cryptographic operations:
\begin{equation}
\label{eq:decomp}
L_i(M_i) = f_i^{(k_i)} \circ f_i^{(k_i - 1)} \circ \cdots \circ f_i^{(1)}(M_i)
\end{equation}
Each operation $f_i^{(j)}$ has a PQC status $\sigma(f_i^{(j)}) \in \statusset$ and a \emph{role} from the set $\{\mathrm{KEX}, \mathrm{AUTH}, \mathrm{ENC}, \mathrm{INT}, \mathrm{KDF}\}$.
\end{definition}

We elaborate on these roles. \textbf{KEX} (key exchange) establishes a shared secret between communicating parties; if this shared secret is recoverable by a quantum adversary, all keys derived from it are also recoverable. 
\textbf{AUTH} (authentication) verifies peer identity via a digital signature or certificate; 
if the signature scheme is quantum-vulnerable, the identity can be forged. \textbf{ENC} (encryption) encrypts the payload using a symmetric or AEAD algorithm. \textbf{INT} (integrity) appends a MAC or AEAD authentication tag.
\textbf{KDF} (key derivation) derives session keys from a shared secret, typically via a hash-based construction like HKDF.

Note that these roles are not independent. 
The security of ENC depends on the security of KEX, because the encryption key is derived from the shared secret produced by the key exchange. 
A \pqsafe{} symmetric cipher offers no protection if the key that feeds it was established through a \pqunsafe{} key exchange. 
We formalize this dependency next.

\subsection{Per-Session and Per-Message Operations}
\label{sec:model:session}

Each active layer's operations divide into two phases that occur at different times and frequencies.

\begin{definition}[Session Establishment]
\label{def:handshake}
The \emph{session establishment} (handshake) at layer $L_i$ is a sequence of operations that produces session key material $K_i$ and an authentication outcome $\mathrm{Auth}_i$:
\begin{equation}
\label{eq:handshake}
L_i^{\mathrm{hs}}: \text{(protocol messages)} \longrightarrow (K_i,\; \mathrm{Auth}_i)
\end{equation}
The PQC status of $K_i$ is determined by the key exchange operation(s) within the handshake. The PQC status of $\mathrm{Auth}_i$ is determined by the authentication operation(s).
\end{definition}

\begin{definition}[Per-Message Data Transformation]
\label{def:data}
After session establishment, each application message is transformed using the established key:
\begin{equation}
\label{eq:data}
L_i^{\mathrm{data}}(M_i) = \mathrm{AEAD}_{K_i}(M_i)
\end{equation}
\end{definition}

The distinction matters because a session 
is established once (or infrequently), while data transformations occur for every message. 
For PQC analysis, the handshake determines the ceiling of what the data phase can achieve: a per-message transformation that uses AES-256-GCM (\pqsafe{} as an algorithm) inherits a \pqunsafe{} effective status if $K_i$ was 
produced by an X25519 key exchange.

We state this as a principle that recurs throughout the paper.

\begin{remark}[Key Material Inheritance]
\label{rem:inheritance}
A \pqsafe{} key derivation function cannot degrade a \pqunsafe{} shared secret into a \pqsafe{} key. If the root of the key derivation chain is \pqunsafe, all derived keys inherit that status regardless of the KDF's own quantum resistance. Concretely, in TLS~1.3, the key derivation chain is:
\[
\text{Random} \xrightarrow{\;\text{X25519}\;} \text{shared\_secret} \xrightarrow{\;\text{HKDF}\;} K_{\mathrm{hs}} \xrightarrow{\;\text{HKDF}\;} K_{\mathrm{app}}
\]
HKDF-SHA384 is \pqsafe, but the root shared secret is \pqunsafe{} (produced by X25519), so $K_{\mathrm{app}}$ is \pqunsafe: a quantum adversary recovers the X25519 shared secret and re-derives every subsequent key.
\end{remark}

\subsection{Effective PQC Status}
\label{sec:model:effective}

We now define the quantity that matters for system-level analysis. The \emph{effective} PQC status of a layer's protection, 
which accounts for the dependency between key exchange and encryption.

\begin{definition}[Effective Confidentiality Status]
\label{def:effconf}
The \emph{effective confidentiality PQC status} of layer $L_i$ is
\begin{equation}
\label{eq:effconf}
\seffconf{L_i} = \min\!\big(\sigma(\mathrm{KEX}_i),\; \sigma(\mathrm{ENC}_i)\big)
\end{equation}
where $\min$ follows the ordering $\classun < \pqunsafe < \pqweak < \pqsafe$ from Definition~\ref{def:status}. If no key exchange is performed (e.g., encryption uses a pre-shared key), $\sigma(\mathrm{KEX}_i)$ is replaced by the PQC status of the key source.
\end{definition}

Intuitively, a layer's confidentiality protection is only as strong as the weaker of its key exchange and its encryption algorithm. 
If the key exchange is \pqunsafe{} but the cipher is \pqsafe, 
a quantum adversary recovers the session key through the key exchange and then decrypts the ciphertext. 
The effective status is \pqunsafe.

\begin{definition}[Effective Authentication Status]
\label{def:effauth}
The \emph{effective authentication PQC status} of layer $L_i$ is
\begin{equation}
\label{eq:effauth}
\seffauth{L_i} = \sigma(\mathrm{AUTH}_i)
\end{equation}
If the layer uses symmetric authentication (e.g., HMAC with a pre-shared key) rather than public-key signatures, $\seffauth{L_i}$ equals the PQC status of the symmetric MAC's key source.
\end{definition}

Authentication status does not depend on the encryption algorithm. It depends only on the signature or MAC scheme used to verify peer identity. 
A layer can have \pqsafe{} confidentiality (via ML-KEM key exchange) and \pqunsafe{} authentication (via ECDSA certificates) simultaneously; 
this is exactly the situation in current hybrid TLS deployments.

Table~\ref{tab:notation} collects the notation introduced in this section.

\begin{table*}[t]
\centering
\footnotesize
\caption{Notation summary.}
\label{tab:notation}
\footnotesize
\begin{tabular}{@{}ll@{}}
\toprule
\textbf{Symbol} & \textbf{Meaning} \\
\midrule
$\statusset$ & PQC status set $\{\classun < \pqunsafe < \pqweak < \pqsafe\}$ \\
$L_i$ & Cryptographic transformation at OSI layer $i$ (sending) \\
$L_i^{-1}$ & Inverse transformation at layer $i$ (receiving) \\
$L_i^{\mathrm{hs}}$ & Session establishment (handshake) at layer $i$ \\
$L_i^{\mathrm{data}}$ & Per-message data transformation at layer $i$ \\
$M_{\mathrm{app}}$ & Application-layer plaintext message \\
$M_{\mathrm{out}}$ & Transmitted frame on the physical medium \\
$f_i^{(j)}$ & The $j$-th cryptographic operation at layer $i$ \\
$\sigma(f)$ & PQC status of operation $f$ \\
$K_i$ & Session key material established at layer $i$ \\
$\layerset$ & Active layer set $(L_{i_1}, \ldots, L_{i_n})$ \\
$\seffconf{L_i}$ & Effective confidentiality PQC status of layer $i$ \\
$\seffauth{L_i}$ & Effective authentication PQC status of layer $i$ \\
$\schainconf$ & Chain-level confidentiality PQC status \\
$\schainauth$ & Chain-level authentication PQC status \\
$\schainmeta$ & Chain-level metadata PQC status \\
$d^*$ & Metadata exposure depth \\
$\dagger$ & Grover-reduced (as opposed to Shor-broken) \\
\bottomrule
\end{tabular}
\end{table*}

\section{Case Studies}
\label{sec:cases}

We apply the formal model from Section~3 
to some commonly occurring scenarios. 
Each case study instantiates the framework with one specific protocol stack and one fixed cryptographic profile, involving a different combination of Layer-2, Layer-3, Layer 5--6, and Layer-7 protections. 
Accordingly, the conclusions of a case study should be understood as properties of the instantiated configuration under analysis, not as universal claims about all devices, operating systems, or deployments of the corresponding protocols. 
Together, these case studies illustrate how the same application data can have radically different quantum vulnerability depending on the protocol stack it traverses.

\subsection{Methodology}
\label{sec:cases:method}

For each scenario, we present three analyses.

The first is a \emph{per-layer cryptographic profile} that records the algorithms at each active layer together with the effective PQC statuses $\seffconf{L_i}$ and $\seffauth{L_i}$, followed by a \emph{segment-by-segment table}. Each segment corresponds to a physical network link between two adjacent nodes (e.g., device~$\to$~access point, or VPN server~$\to$~web server). For each segment, the table shows which encryption layers are active on that link, the composite PQC status an adversary on that segment would face, and what specific data is exposed at that point if encryptions are stripped.

The second is an \emph{endpoint vulnerability posture} table. Where the segment table describes what is on the wire between two nodes, the endpoint table describes what is exposed \emph{at} each node---including the sender and final recipient. For each endpoint, we separate three concerns: \emph{classical exposure} (what the node sees today, by design, without any cryptographic attack), \emph{HNDL exposure} (what an adversary who captures traffic at this point could additionally recover with a future CRQC), and \emph{quantum-resistant protection} (which layers, if any, remain unbreakable even by a quantum adversary). This separation makes explicit whether quantum computing creates genuinely new exposure or merely extends existing classical exposure to new adversaries.

The third is a \emph{quantum exposure analysis} that traces what an HNDL adversary learns as she peels successive encryption layers from a captured frame. Each row corresponds to a \emph{depth}~$d$: depth~0 is the wire observation, depth~1 is the result of breaking the outermost layer, and so on. The HNDL column records whether that depth is reachable and what data is harvestable.

\begin{definition}[HNDL Exposure Depth]
\label{def:hndl-depth}
For a message transformation chain with $n$ active layers ordered outermost to innermost, the \emph{HNDL exposure depth} is
\begin{equation}
\label{eq:hndl-depth}
d^* = \max\{d : \seffconf{L_{i_k}} \neq \pqsafe \text{ for all } k \leq d\}
\end{equation}
All data at depths $1, \ldots, d^*$ is harvestable. If $d^* = n$, the application plaintext itself is harvestable.
\end{definition}

We use the notation $\pqunsafe^\dagger$ from Section~3 (Remark~2) to distinguish Grover-reduced vulnerabilities (AES-128 at 64-bit effective) from Shor-broken ones (ECDH, RSA).

\subsection{Case Study 1: iOS-to-iOS iMessage}
\label{sec:cs1}

\noindent\textbf{Configuration.} We consider an instantiated communication stack in which two iPhones running iOS~18 exchange iMessages over a WPA3-Personal Wi-Fi network. 
For this case study, the stack is taken to have three active cryptographic layers: WPA3-SAE at Layer~2, TLS~1.3 to Apple's iMessage relay at Layers~5--6, and Apple's PQ3~\cite{stebila2024} end-to-end protection at Layer~7. 
The per-layer cryptographic profile analyzed below is the fixed profile for this instantiated configuration.

\smallskip
\noindent\textbf{Active layers:} $\layerset = (L_2^{\text{WPA3}},\; L_{5\text{-}6}^{\text{TLS}},\; L_7^{\text{PQ3}})$, \; $n = 3$.

\smallskip
\noindent\textbf{Sending chain:}
$M_{\mathrm{out}} = L_2^{\text{WPA3}}\!\Big(L_{5\text{-}6}^{\text{TLS}}\!\big(L_7^{\text{PQ3}}(M_{\mathrm{app}})\big)\Big)$

\subsubsection{Per-Layer Cryptographic Profile}

\begin{table*}[t]
\centering
\caption{CS1: Per-layer cryptographic profile.}
\label{tab:cs1-profile}
\footnotesize
\begin{tabular}{@{}llllcc@{}}
\toprule
\textbf{Layer} & \textbf{Protocol} & \textbf{KEX} & \textbf{AUTH} & $\boldsymbol{\sigma_{\mathrm{conf}}}$ & $\boldsymbol{\sigma_{\mathrm{auth}}}$ \\
\midrule
L2 & WPA3-SAE & EC Dragonfly (P-256 class) & SAE implicit (EC) & \pqunsafe & \pqunsafe \\
L5--6 & TLS 1.3 & X25519 & ECDSA-P256 cert & \pqunsafe & \pqunsafe \\
L7 & PQ3 & Hybrid ECC+PQ initial establishment and ratcheting & ECDSA-P256 signing & \pqsafe & \pqunsafe \\
\bottomrule
\multicolumn{6}{@{}L{\dimexpr\textwidth-2\tabcolsep}@{}}{\footnotesize
Data encryption: L2 = AES-128-CCMP (\pqunsafe$^\dagger$), L5--6 = AES-256-GCM (\pqsafe), L7 = AES-256-CTR (\pqsafe). 
For PQ3, initial session establishment combines a Kyber-1024 KEM with P-256 ECDH, and ongoing rekeying combines a per-message P-256 ECDH ratchet with a periodic Kyber-768 ratchet.}
\end{tabular}
\end{table*}

In our abstraction, PQ3 authentication is classified as \pqunsafe{} because its per-message signature primitive is ECDSA-P256. 
This classification abstracts away additional platform-level hardening and verification mechanisms and focuses only on the PQC status of the underlying authentication primitive.

\subsubsection{Segment-by-Segment Analysis}

Table~\ref{tab:cs1-seg} traces the message across each physical network link. On each segment, the ``Active Layers'' column lists the encryption layers whose protection covers that link. The composite $\sigma_{\mathrm{conf}}$ for the segment is $\max$ of the per-layer confidentiality statuses (one \pqsafe{} layer suffices). The ``Exposed at This Node'' column describes what an entity \emph{at the termination point} of a stripped layer can see---even without quantum capabilities.

\begin{table*}[t]
\centering
\caption{CS1: Segment-by-segment analysis (sender's outbound path).}
\label{tab:cs1-seg}
\footnotesize
\begin{tabular}{@{} L{2.2cm} L{2.3cm} cc L{9cm} @{}}
\toprule
\textbf{Segment} & \textbf{Active Layers} & $\boldsymbol{\sigma_{\mathrm{conf}}}$ & $\boldsymbol{\sigma_{\mathrm{auth}}}$ & \textbf{Exposed at Receiving Node} \\
\midrule
iPhone~A $\to$ AP & {L2} + {L5--6} + {L7} & \pqsafe & \pqunsafe & \textbf{AP strips L2.} Sees: outer IP headers (device IP $\to$ Apple relay IP), TCP port 443, TLS record headers. Cannot see TLS payload or PQ3-protected content. \\
\addlinespace
AP $\to$ Apple Relay & {L5--6} + {L7} & \pqsafe & \pqunsafe & \textbf{Relay strips L5--6.} Sees: PQ3-protected ciphertext blobs and relay-visible messaging metadata (e.g., timing and ciphertext size information). Does not obtain message plaintext. \\
\addlinespace
Apple Relay $\to$ AP & {L5--6$'$} + {L7} & \pqsafe & \pqunsafe & New TLS session to iPhone~B. AP strips L2 on final hop. Same relay-visible metadata exposure as above. \\
\addlinespace
AP $\to$ iPhone~B & {L2$'$} + {L5--6$'$} + {L7} & \pqsafe & \pqunsafe & \textbf{iPhone~B strips all.} Recovers $M_{\mathrm{app}}$. \\
\bottomrule
\multicolumn{5}{@{}L{\dimexpr\textwidth-2\tabcolsep}@{}}{\footnotesize
Primed layers ($'$) indicate new sessions on the relay-to-recipient path. PQ3 is not terminated at the relay; the relay forwards PQ3-protected ciphertext and does not obtain the message plaintext.}
\end{tabular}
\end{table*}

The relay server is a \emph{trust boundary} where TLS protection disappears. Even without a quantum computer, Apple's relay infrastructure has access to relay-visible messaging metadata. PQ3 ensures that this exposure does not extend to message content.

\subsubsection{Endpoint Vulnerability Posture}

Table~\ref{tab:cs1-endpoint} shows the vulnerability posture at each network node. For each endpoint, we separate what the node sees \emph{classically} (today, by design) from what an HNDL adversary who captures traffic at that point could \emph{additionally} recover with a future quantum computer. The ``Quantum-Resistant'' column identifies what remains protected even against a CRQC-equipped adversary.

\begin{table*}[t]
\centering
\caption{CS1: Endpoint vulnerability posture.}
\label{tab:cs1-endpoint}
\footnotesize
\begin{tabular}{@{} L{1.5cm} L{1.8cm} L{5.5cm} L{5.5cm} L{2.3cm} @{}}
\toprule
\textbf{Endpoint} & \textbf{Layers Remaining} & \textbf{Classical Exposure} & \textbf{HNDL Exposure (quantum)} & \textbf{Quantum-Resistant} \\
\midrule
iPhone~A (sender) & L2+L5--6+L7 (pre-tx) & Full plaintext $M_{\mathrm{app}}$ (origin device) & N/A --- data not yet transmitted & All outbound layers \\
\addlinespace
AP (sender) & L5--6 + L7 & IP headers: device IP $\to$ relay IP; TCP:443; TLS record headers & Break L5--6 (\pqunsafe, Shor) $\to$ relay-visible messaging metadata. \textbf{Blocked} at L7 (PQ3). & L7: PQ3 hybrid ECC+PQ confidentiality \\
\addlinespace
Apple Relay & L7 only & PQ3-protected ciphertext and relay-visible messaging metadata & PQ3-protected ciphertext only. \textbf{No additional content recovery.} & L7: PQ3 hybrid ECC+PQ confidentiality \\
\addlinespace
AP (recip.) & L5--6$'$ + L7 & Same as sender AP (symmetric) & Same as sender AP & L7: PQ3 hybrid ECC+PQ confidentiality \\
\addlinespace
iPhone~B (recip.) & None & Full plaintext $M_{\mathrm{app}}$ (destination device) & N/A (endpoint) & --- \\
\bottomrule
\end{tabular}
\end{table*}

The endpoint table reveals that the classical and quantum exposure boundaries \emph{converge} at the relay. Apple's relay server sees relay-visible messaging metadata today, by design, without any cryptographic attack. An HNDL adversary who captures traffic anywhere upstream of the relay and later obtains a CRQC recovers the same class of metadata---but nothing more, because PQ3 blocks at every endpoint. The quantum threat does not create message-content exposure at the relay; it extends relay-visible metadata exposure to adversaries who are not Apple.

\subsubsection{Quantum Exposure}

\begin{table*}[t]
\centering
\caption{CS1: HNDL quantum exposure analysis.}
\label{tab:cs1-exposure}
\footnotesize
\begin{tabular}{@{}cll L{6.0cm} c L{6cm}@{}}
\toprule
$\boldsymbol{d}$ & \textbf{Layer} & $\boldsymbol{\sigma_{\mathrm{conf}}}$ & \textbf{Newly Revealed Data} & \textbf{HNDL} & \textbf{Harvestable Data} \\
\midrule
0 & \textit{Wire} & --- & 802.11 headers: src/dst MAC, BSS~ID, frame type & --- & Device presence and activity timing \\
\addlinespace
1 & L2: WPA3 & \pqunsafe & IP headers: device IP, Apple relay IP; TCP port 443; TLS record headers & Yes & Communication pattern to Apple servers \\
\addlinespace
2 & L5--6: TLS & \pqunsafe & Relay-visible messaging metadata; PQ3-protected ciphertext & Yes & Messaging metadata visible after TLS removal \\
\addlinespace
3 & L7: PQ3 & \pqsafe & \textbf{BLOCKED} --- PQ3's end-to-end confidentiality remains protected by its hybrid ECC+PQ design and ongoing post-quantum ratcheting & No & Content is \emph{not} harvestable \\
\bottomrule
\end{tabular}
\end{table*}

\noindent\textbf{Chain composition:}
\begin{align}
\schainconf &= \max(\pqunsafe,\; \pqunsafe,\; \pqsafe) = \pqsafe  \label{eq:cs1-conf} \\
\schainauth &= \min(\pqunsafe,\; \pqunsafe,\; \pqunsafe) = \pqunsafe  \label{eq:cs1-auth} \\
\schainmeta &= \seffconf{L_2} = \pqunsafe  \label{eq:cs1-meta} \\
d^* &= 2 \notag
\end{align}

The iMessage scenario demonstrates the central asymmetry of this paper. A single \pqsafe{} layer at Layer~7 makes the chain \pqsafe{} for payload confidentiality, but authentication remains \pqunsafe{} at every layer and metadata is exposed to depth~$d^* = 2$. An HNDL adversary can eventually recover relay-visible messaging metadata but cannot recover message content. The segment analysis adds a further insight: the relay server is a classical trust boundary that mirrors the quantum exposure boundary. Apple can see today, without any quantum computer, the same class of relay-visible metadata that an HNDL adversary could recover in a quantum future.
\subsection{Case Study 2: Linux-to-Linux HTTPS over WPA2-PSK}
\label{sec:cs2}

\noindent\textbf{Configuration.} We consider an instantiated communication stack in which two Linux machines (Ubuntu 24.04) communicate via HTTPS over a WPA2-Personal (PSK) Wi-Fi network, with no VPN and no application-layer encryption. 
The per-layer cryptographic profile analyzed below is the fixed profile for this instantiated configuration.

\smallskip
\noindent\textbf{Active layers:} $\layerset = (L_2^{\text{WPA2-PSK}},\; L_{5\text{-}6}^{\text{TLS}})$, \; $n = 2$.

\smallskip
\noindent\textbf{Sending chain:}
$M_{\mathrm{out}} = L_2^{\text{WPA2-PSK}}\!\Big(L_{5\text{-}6}^{\text{TLS}}(M_{\mathrm{app}})\Big)$

\subsubsection{Per-Layer Cryptographic Profile}

\begin{table*}[t]
\centering
\caption{CS2: Per-layer cryptographic profile.}
\label{tab:cs2-profile}
\footnotesize
\begin{tabular}{@{}llllcc@{}}
\toprule
\textbf{Layer} & \textbf{Protocol} & \textbf{KEX} & \textbf{AUTH} & $\boldsymbol{\sigma_{\mathrm{conf}}}$ & $\boldsymbol{\sigma_{\mathrm{auth}}}$ \\
\midrule
L2 & WPA2-PSK & None (symmetric PSK) & HMAC-SHA1 MIC & \pqunsafe$^\dagger$ & \pqweak \\
L5--6 & TLS 1.3 & X25519 & ECDSA-P256 cert & \pqunsafe & \pqunsafe \\
\bottomrule
\multicolumn{6}{@{}L{\dimexpr\textwidth-2\tabcolsep}@{}}{\footnotesize Data encryption: L2 = AES-128-CCMP (\pqunsafe$^\dagger$ under Grover), L5--6 = AES-256-GCM (\pqsafe). L2 uses no public-key KEX; $\sigma_{\mathrm{conf}} = \min(\pqweak^{\text{key}}, \pqunsafe^{\dagger\text{AES-128}}) = \pqunsafe^\dagger$.}
\end{tabular}
\end{table*}

\subsubsection{Segment-by-Segment Analysis}

\begin{table*}[t]
\centering
\caption{CS2: Segment-by-segment analysis.}
\label{tab:cs2-seg}
\footnotesize
\begin{tabular}{@{} L{2.2cm} L{2.3cm} cc L{9cm} @{}}
\toprule
\textbf{Segment} & \textbf{Active Layers} & $\boldsymbol{\sigma_{\mathrm{conf}}}$ & $\boldsymbol{\sigma_{\mathrm{auth}}}$ & \textbf{Exposed at Receiving Node} \\
\midrule
Linux~A $\to$ AP & {L2} + {L5--6} & \pqunsafe & \pqunsafe & \textbf{AP strips L2.} Sees: IP headers (client IP, server IP), TCP port 443, TLS ClientHello with SNI (server hostname), TLS record sizes. \\
\addlinespace
AP $\to$ Server & {L5--6} only & \pqunsafe & \pqunsafe & \textbf{Server strips L5--6.} Recovers full HTTP content: $M_{\mathrm{app}}$. On this segment, an eavesdropper sees TLS ciphertext with no WPA2 wrapper. \\
\bottomrule
\end{tabular}
\end{table*}

The segment table highlights that the wired segment (AP~$\to$~Server) has \emph{only} TLS protecting it. An adversary with physical access to the wired network (or any router between AP and server) faces one fewer encryption layer than a wireless eavesdropper. For HNDL purposes, both segments are equally vulnerable (both have $\sigma_{\mathrm{conf}} = \pqunsafe$), but the wired adversary's task is simpler: break TLS only, rather than WPA2 + TLS.

\subsubsection{Endpoint Vulnerability Posture}

\begin{table*}[t]
\centering
\caption{CS2: Endpoint vulnerability posture.}
\label{tab:cs2-endpoint}
\footnotesize
\begin{tabular}{@{} L{1.5cm} L{1.8cm} L{5.5cm} L{5.5cm} L{2.3cm} @{}}
\toprule
\textbf{Endpoint} & \textbf{Layers Remaining} & \textbf{Classical Exposure} & \textbf{HNDL Exposure (quantum)} & \textbf{Quantum-Resistant} \\
\midrule
Linux~A (sender) & L2+L5--6 (pre-tx) & Full plaintext $M_{\mathrm{app}}$ (origin) & N/A (not yet transmitted) & All outbound layers \\
\addlinespace
AP & L5--6 only & IP headers: client/server IPs; TCP:443; TLS ClientHello SNI; TLS record sizes & Break L5--6 (\pqunsafe, Shor on X25519) $\to$ \textbf{full HTTP content}. No blocking layer. & \textbf{None} --- no \pqsafe{} layer remains \\
\addlinespace
Server & None & Full plaintext $M_{\mathrm{app}}$ (destination) & N/A (endpoint) & --- \\
\bottomrule
\end{tabular}
\end{table*}

The critical entry is the AP. After stripping WPA2, only TLS remains---and TLS is \pqunsafe. An HNDL adversary who captures traffic at the AP (or on the wired network beyond it) faces a single Shor attack on X25519 to recover all application data. No \pqsafe{} layer blocks further decryption. Contrast this with CS1, where PQ3 blocks at every endpoint; here, there is no backstop.

\subsubsection{Quantum Exposure}

\begin{table*}[t]
\centering
\caption{CS2: HNDL quantum exposure analysis.}
\label{tab:cs2-exposure}
\footnotesize
\begin{tabular}{@{}cll L{6cm} c L{6cm}@{}}
\toprule
$\boldsymbol{d}$ & \textbf{Layer} & $\boldsymbol{\sigma_{\mathrm{conf}}}$ & \textbf{Newly Revealed Data} & \textbf{HNDL} & \textbf{Harvestable Data} \\
\midrule
0 & \textit{Wire} & --- & 802.11 headers: MAC addrs, BSS~ID, frame sizes & --- & Device presence, activity timing \\
\addlinespace
1 & L2: WPA2-PSK & \pqunsafe$^\dagger$ & IP headers: src/dst IPs; TCP port 443; TLS SNI (hostname); TLS record sizes and timing & Yes$^\dagger$ & Server identities; traffic volumes per destination \\
\addlinespace
2 & L5--6: TLS 1.3 & \pqunsafe & Full HTTP content: URLs, cookies, auth tokens, form data, API bodies & Yes & \textbf{All application data} \\
\bottomrule
\end{tabular}
\end{table*}

\noindent\textbf{Chain composition:}
\begin{align}
\schainconf &= \max(\pqunsafe^\dagger,\; \pqunsafe) = \pqunsafe \label{eq:cs2-conf} \\
\schainauth &= \min(\pqweak,\; \pqunsafe) = \pqunsafe \label{eq:cs2-auth} \\
\schainmeta &= \seffconf{L_2} = \pqunsafe^\dagger \label{eq:cs2-meta} \\
d^* &= 2 = n \notag
\end{align}

This scenario is fully \pqunsafe: an HNDL adversary can eventually recover all application data. However, the mechanisms differ by layer. Layer~2's vulnerability is Grover-based (AES-128 reduced to 64-bit effective), fixable by upgrading to AES-256. Layer 5--6's vulnerability is Shor-based (X25519), requiring protocol replacement. WPA2-PSK's authentication ($\pqweak$) is the only non-\pqunsafe{} authentication in any scenario in this study---because it uses only symmetric primitives.

% CASE STUDY 3

\subsection{Case Study 3: HTTPS over WPA2-Enterprise}
\label{sec:cs3}

\noindent\textbf{Configuration.} We consider an instantiated communication stack in which a corporate laptop communicates via HTTPS over WPA2-Enterprise (EAP-TLS), with certificate-based mutual authentication at Layer~2, TLS~1.3 at Layers~5--6, and no VPN or application-layer encryption. 
The case study analyzes one fixed cryptographic profile for this enterprise-style configuration rather than all possible WPA2-Enterprise or TLS deployments.

\smallskip
\noindent\textbf{Active layers:} $\layerset = (L_2^{\text{WPA2-Ent}},\; L_{5\text{-}6}^{\text{TLS}})$, \; $n = 2$.

\smallskip
\noindent\textbf{Sending chain:}
$M_{\mathrm{out}} = L_2^{\text{WPA2-Ent}}\!\Big(L_{5\text{-}6}^{\text{TLS}}(M_{\mathrm{app}})\Big)$

\subsubsection{Per-Layer Cryptographic Profile}

\begin{table*}[t]
\centering
\caption{CS3: Per-layer cryptographic profile.}
\label{tab:cs3-profile}
\footnotesize
\begin{tabular}{@{}llllcc@{}}
\toprule
\textbf{Layer} & \textbf{Protocol} & \textbf{KEX} & \textbf{AUTH} & $\boldsymbol{\sigma_{\mathrm{conf}}}$ & $\boldsymbol{\sigma_{\mathrm{auth}}}$ \\
\midrule
L2 & WPA2-Ent & ECDHE-P256 (EAP-TLS) & RSA-2048/ECDSA certs & \pqunsafe & \pqunsafe \\
L5--6 & TLS 1.3 & X25519 & ECDSA-P256 cert & \pqunsafe & \pqunsafe \\
\bottomrule
\multicolumn{6}{@{}L{\dimexpr\textwidth-2\tabcolsep}@{}}{\footnotesize Data encryption: L2 = AES-128-CCMP (\pqunsafe$^\dagger$), L5--6 = AES-256-GCM (\pqsafe). L2 $\sigma_{\mathrm{conf}} = \min(\pqunsafe^{\text{ECDHE}}, \pqunsafe^{\dagger\text{AES-128}}) = \pqunsafe$; Shor on ECDHE dominates.}
\end{tabular}
\end{table*}

\subsubsection{Segment-by-Segment Analysis}

\begin{table*}[t]
\centering
\caption{CS3: Segment-by-segment analysis.}
\label{tab:cs3-seg}
\footnotesize
\begin{tabular}{@{} L{2.2cm} L{2.3cm} cc L{9cm} @{}}
\toprule
\textbf{Segment} & \textbf{Active Layers} & $\boldsymbol{\sigma_{\mathrm{conf}}}$ & $\boldsymbol{\sigma_{\mathrm{auth}}}$ & \textbf{Exposed at Receiving Node} \\
\midrule
Laptop $\to$ AP & {L2} + {L5--6} & \pqunsafe & \pqunsafe & \textbf{AP strips L2.} Sees: IP headers, destination server IP, TCP port 443, TLS records. Also: EAP-TLS auth exchange (client/server certs, ECDHE params) was visible during association. \\
\addlinespace
AP $\to$ Server & {L5--6} only & \pqunsafe & \pqunsafe & \textbf{Server strips L5--6.} Recovers $M_{\mathrm{app}}$. No L2 protection on this segment. \\
\bottomrule
\multicolumn{5}{@{}L{\dimexpr\textwidth-2\tabcolsep}@{}}{\footnotesize During 802.1X association, EAP identity frames (containing employee usernames) are sent in cleartext before encryption is established.}
\end{tabular}
\end{table*}

\subsubsection{Endpoint Vulnerability Posture}

\begin{table*}[t]
\centering
\caption{CS3: Endpoint vulnerability posture.}
\label{tab:cs3-endpoint}
\footnotesize
\begin{tabular}{@{} L{1.5cm} L{1.6cm} L{5.5cm} L{5.5cm} L{2.3cm} @{}}
\toprule
\textbf{Endpoint} & \textbf{Layers Remaining} & \textbf{Classical Exposure} & \textbf{HNDL Exposure (quantum)} & \textbf{Quantum-Resistant} \\
\midrule
Laptop (sender) & L2+L5--6 (pre-tx) & Full plaintext $M_{\mathrm{app}}$ (origin) & N/A (not yet transmitted) & All outbound layers \\
\addlinespace
AP & L5--6 only & IP headers; TLS records; during association: EAP-TLS certs and ECDHE params (employee cert DNs visible) & Break L5--6 (\pqunsafe, Shor) $\to$ \textbf{full HTTP content}. No blocking layer. & \textbf{None}( no \pqsafe{} layer remains) \\
\addlinespace
RADIUS & L5--6 (not on data path) & EAP-TLS auth exchange: client/server certs, derives PMK. Does \emph{not} see data-phase traffic. & N/A for data traffic (auth-only role) & --- \\
\addlinespace
Web Server & None & Full plaintext $M_{\mathrm{app}}$ (destination) & N/A (endpoint) & --- \\
\bottomrule
\end{tabular}
\end{table*}

The AP's row is the most concerning. As in CS2, after stripping Layer~2 the AP exposes TLS-only traffic with no \pqsafe{} backstop. But the mechanism is worse than CS2: the WPA2-Enterprise AP has richer classical exposure (employee certificate DNs from the EAP-TLS exchange, cleartext EAP identity frames) and its Layer-2 vulnerability is Shor-based rather than Grover-based, meaning it cannot be fixed by a cipher upgrade alone. The RADIUS server, though it participates in authentication, is not on the data path and does not see post-handshake application traffic.

\subsubsection{Quantum Exposure}

\begin{table*}[t]
\centering
\caption{CS3: HNDL quantum exposure analysis.}
\label{tab:cs3-exposure}
\footnotesize
\begin{tabular}{@{}cll L{6cm} c L{6cm}@{}}
\toprule
$\boldsymbol{d}$ & \textbf{Layer} & $\boldsymbol{\sigma_{\mathrm{conf}}}$ & \textbf{Newly Revealed Data} & \textbf{HNDL} & \textbf{Harvestable Data} \\
\midrule
0 & \textit{Wire} & --- & 802.11 headers; EAP identity (employee username in cleartext) & --- & Employee usernames; corporate SSID; device MACs \\
\addlinespace
1 & L2: WPA2-Ent & \pqunsafe & IP headers; TLS records; additionally: EAP-TLS handshake (client/server certs, ECDHE parameters) & Yes & Internal topology; server IPs; employee certificate DNs \\
\addlinespace
2 & L5--6: TLS 1.3 & \pqunsafe & Full HTTP: internal apps, API calls, documents, session tokens & Yes & \textbf{All corporate data} \\
\bottomrule
\end{tabular}
\end{table*}

\noindent\textbf{Chain composition:}
\begin{align}
\schainconf &= \max(\pqunsafe,\; \pqunsafe) = \pqunsafe \label{eq:cs3-conf} \\
\schainauth &= \min(\pqunsafe,\; \pqunsafe) = \pqunsafe \label{eq:cs3-auth} \\
\schainmeta &= \seffconf{L_2} = \pqunsafe \label{eq:cs3-meta} \\
d^* &= 2 = n \notag
\end{align}

\subsubsection{Direct Comparison: WPA2-Personal vs.\ WPA2-Enterprise}

Table~\ref{tab:cs2v3} isolates the Layer-2 difference. Both scenarios use identical TLS~1.3 and identical AES-128-CCMP data encryption. They differ only in key establishment.

\begin{table*}[t]
\centering
\caption{Layer-2 comparison: WPA2-Personal vs.\ WPA2-Enterprise.}
\label{tab:cs2v3}
\footnotesize
\begin{tabular}{@{}lccll@{}}
\toprule
& \multicolumn{2}{c}{\textbf{PQC Status}} & \multicolumn{2}{c}{\textbf{Vulnerability Mechanism}} \\
\cmidrule(lr){2-3} \cmidrule(lr){4-5}
\textbf{L2 Variant} & $\sigma_{\mathrm{conf}}$ & $\sigma_{\mathrm{auth}}$ & Conf & Auth \\
\midrule
WPA2-PSK & \pqunsafe$^\dagger$ & \pqweak & Grover (AES-128) & Grover (sym.\ MAC) \\
WPA2-Ent & \pqunsafe & \pqunsafe & Shor (ECDHE) & Shor (RSA/ECDSA) \\
\midrule
\multicolumn{3}{@{}l}{\textbf{Classical security}} & \multicolumn{2}{l}{Enterprise $\gg$ Personal} \\
\multicolumn{3}{@{}l}{\textbf{Quantum security}} & \multicolumn{2}{l}{Personal $>$ Enterprise} \\
\multicolumn{3}{@{}l}{\textbf{Remediation cost}} & \multicolumn{2}{l}{Personal: config $\ll$ Enterprise: protocol} \\
\bottomrule
\end{tabular}
\end{table*}

WPA2-Enterprise is classically superior by every measure. Yet its quantum posture is strictly worse: Shor-broken ECDHE versus Grover-reduced AES-128. Worse still, the remediation paths diverge sharply. Upgrading WPA2-PSK to AES-256 is a cipher-suite configuration change. Upgrading WPA2-Enterprise requires replacing the entire EAP-TLS key exchange with a post-quantum mechanism, touching certificate infrastructure, RADIUS servers, and supplicant software. This is a concrete, quantified instance of the classical--quantum security tension in RC6.

% CASE STUDY 4

\subsection{Case Study 4: HTTPS over WPA3 with WireGuard VPN}
\label{sec:cs4}

\noindent\textbf{Configuration.} We consider an instantiated communication stack in which a user on a WPA3-Personal Wi-Fi network routes all traffic through a WireGuard VPN and then connects to an HTTPS website via TLS~1.3. 
For this case study, the stack is taken to have three active cryptographic layers spanning Layers~2, 3, and 5--6. 
The per-layer cryptographic profile analyzed below is the fixed profile for this instantiated configuration.

\smallskip
\noindent\textbf{Active layers:} $\layerset = (L_2^{\text{WPA3}},\; L_3^{\text{WG}},\; L_{5\text{-}6}^{\text{TLS}})$, \; $n = 3$.

\smallskip
\noindent\textbf{Sending chain:}
$M_{\mathrm{out}} = L_2^{\text{WPA3}}\!\Big(L_3^{\text{WG}}\!\big(L_{5\text{-}6}^{\text{TLS}}(M_{\mathrm{app}})\big)\Big)$

WireGuard uses Noise~IK~\cite{noise}: two Curve25519 DH operations for key exchange, Curve25519 static keys for authentication, and ChaCha20-Poly1305 for data encryption.

\subsubsection{Per-Layer Cryptographic Profile}

\begin{table*}[t]
\centering
\caption{CS4: Per-layer cryptographic profile.}
\label{tab:cs4-profile}
\footnotesize
\begin{tabular}{@{}llllcc@{}}
\toprule
\textbf{Layer} & \textbf{Protocol} & \textbf{KEX} & \textbf{AUTH} & $\boldsymbol{\sigma_{\mathrm{conf}}}$ & $\boldsymbol{\sigma_{\mathrm{auth}}}$ \\
\midrule
L2 & WPA3-SAE & EC Dragonfly (P-256) & SAE implicit (EC) & \pqunsafe & \pqunsafe \\
L3 & WireGuard & Curve25519 DH & Curve25519 static & \pqunsafe & \pqunsafe \\
L5--6 & TLS 1.3 & X25519 & ECDSA-P256 cert & \pqunsafe & \pqunsafe \\
\bottomrule
\multicolumn{6}{@{}L{\dimexpr\textwidth-2\tabcolsep}@{}}{\footnotesize Data encryption: L2 = AES-128-CCMP (\pqunsafe$^\dagger$), L3 = ChaCha20-Poly1305 (\pqsafe), L5--6 = AES-256-GCM (\pqsafe). WireGuard optionally supports a PSK mixed into the Noise handshake; with a Q-Safe PSK, $\seffconf{L_3}$ becomes \pqsafe. Standard deployments do not use this option.}
\end{tabular}
\end{table*}

\subsubsection{Segment-by-Segment Analysis}

\begin{table*}[t]
\centering
\caption{CS4: Segment-by-segment analysis.}
\label{tab:cs4-seg}
\footnotesize
\begin{tabular}{@{} L{2.2cm} L{2.7cm} cc L{9cm} @{}}
\toprule
\textbf{Segment} & \textbf{Active Layers} & $\boldsymbol{\sigma_{\mathrm{conf}}}$ & $\boldsymbol{\sigma_{\mathrm{auth}}}$ & \textbf{Exposed at Receiving Node} \\
\midrule
Device $\to$ AP & {L2} + {L3} + {L5--6} & \pqunsafe & \pqunsafe & \textbf{AP strips L2.} Sees: outer IP (device IP $\to$ VPN server IP), UDP port 51820, WireGuard handshake/data. Knows VPN is in use; cannot see inner traffic. \\
\addlinespace
AP $\to$ VPN & {L3} + {L5--6} & \pqunsafe & \pqunsafe & \textbf{VPN strips L3.} Sees: inner IP (client virtual IP $\to$ web server IP), TCP port 443, TLS ClientHello with SNI. \textbf{VPN sees browsing destinations.} \\
\addlinespace
VPN $\to$ Web & {L5--6} only & \pqunsafe & \pqunsafe & \textbf{Web server strips L5--6.} Recovers $M_{\mathrm{app}}$. On this segment, traffic appears as ordinary HTTPS---no VPN wrapper. \\
\bottomrule
\end{tabular}
\end{table*}

The segment table makes explicit that the VPN server is a full trust boundary for browsing metadata. After stripping WireGuard, the VPN server sees exactly which websites the user visits, when, and how much data is exchanged. This is true today, classically. In a quantum future, an HNDL adversary who captured traffic on the wireless segment (Device~$\to$~AP) would recover the same information by breaking WPA3 and then WireGuard---reaching the same metadata the VPN server already sees.

\subsubsection{Endpoint Vulnerability Posture}

\begin{table*}[t]
\centering
\caption{CS4: Endpoint vulnerability posture.}
\label{tab:cs4-endpoint}
\footnotesize
\begin{tabular}{@{} L{1.5cm} L{1.8cm} L{5.5cm} L{5.5cm} L{2.3cm} @{}}
\toprule
\textbf{Endpoint} & \textbf{Layers Remaining} & \textbf{Classical Exposure} & \textbf{HNDL Exposure (quantum)} & \textbf{Quantum-Resistant} \\
\midrule
Device (sender) & L2+L3+L5--6 (pre-tx) & Full plaintext $M_{\mathrm{app}}$ (origin) & N/A --- not yet transmitted & All outbound layers \\
\addlinespace
AP & L3+L5--6 & Outer IP (device $\to$ VPN server); UDP:51820; WireGuard handshake/data & Break L3 (\pqunsafe, Shor) $\to$ inner IP, TLS SNI. Break L5--6 (\pqunsafe) $\to$ \textbf{all HTTP content}. & \textbf{None} \\
\addlinespace
VPN Server & L5--6 only & Inner IP: virtual IP $\to$ web server; TCP:443; TLS SNI (hostname); traffic volumes. \textbf{Browsing destinations visible.} & Break L5--6 (\pqunsafe, Shor) $\to$ \textbf{full HTTP content}. No blocking layer. & \textbf{None} \\
\addlinespace
Web Server & None & Full plaintext $M_{\mathrm{app}}$ (destination) & N/A (endpoint) & --- \\
\bottomrule
\end{tabular}
\end{table*}

The endpoint table makes the VPN's failure starkly visible. At the VPN server, the classical and HNDL columns tell related but distinct stories. \emph{Classically}, the VPN server sees browsing destinations (which websites, when, how much data) but not content---TLS still protects the payload. \emph{Under HNDL}, even TLS fails: a quantum adversary who captures traffic at the VPN server recovers everything, because no \pqsafe{} layer remains. The ``Quantum-Resistant'' column reads \textbf{None} at every intermediate node---three layers of encryption and not a single quantum backstop.

Compare with CS1's endpoint table, where every intermediate node shows ``L7: PQ3 (Kyber-1024)'' in the Quantum-Resistant column. That single end-to-end \pqsafe{} layer transforms every endpoint from fully exposed to content-protected.

\subsubsection{Quantum Exposure}

\begin{table*}[t]
\centering
\caption{CS4: HNDL quantum exposure analysis.}
\label{tab:cs4-exposure}
\footnotesize
\begin{tabular}{@{}cll L{6.0cm} c L{6cm}@{}}
\toprule
$\boldsymbol{d}$ & \textbf{Layer} & $\boldsymbol{\sigma_{\mathrm{conf}}}$ & \textbf{Newly Revealed Data} & \textbf{HNDL} & \textbf{Harvestable Data} \\
\midrule
0 & \textit{Wire} & --- & 802.11 headers: MACs, BSS~ID, payload sizes & --- & Device presence and timing \\
\addlinespace
1 & L2: WPA3 & \pqunsafe & Outer IP: device $\to$ VPN server; UDP 51820; WireGuard fingerprint & Yes & \textbf{VPN usage exposed}; VPN server identity \\
\addlinespace
2 & L3: WireGuard & \pqunsafe & Inner IP: virtual IP $\to$ web server; TCP 443; TLS SNI (hostname) & Yes & \textbf{Browsing destinations}; defeats VPN privacy \\
\addlinespace
3 & L5--6: TLS 1.3 & \pqunsafe & Full HTTP content: URLs, cookies, auth tokens, response bodies & Yes & \textbf{All application data} \\
\bottomrule
\end{tabular}
\end{table*}

\noindent\textbf{Chain composition:}
\begin{align}
\schainconf &= \max(\pqunsafe,\; \pqunsafe,\; \pqunsafe) = \pqunsafe \label{eq:cs4-conf} \\
\schainauth &= \min(\pqunsafe,\; \pqunsafe,\; \pqunsafe) = \pqunsafe \label{eq:cs4-auth} \\
\schainmeta &= \seffconf{L_2} = \pqunsafe \label{eq:cs4-meta} \\
d^* &= 3 = n \notag
\end{align}

This is the worst case in our study. Three layers of encryption provide zero quantum protection. The HNDL depth equals the total layer count ($d^* = n = 3$), meaning the application plaintext itself is harvestable. An adversary who records the triply-encrypted traffic today can, upon obtaining a CRQC, recover the complete web browsing session.

The VPN is especially counterproductive. Its core value proposition (hiding browsing destinations from the local network and ISP) is defeated at depth~2, where the adversary recovers the inner IP headers and TLS SNI that WireGuard was supposed to conceal. The additional protocol complexity and latency of the VPN tunnel deliver no quantum security benefit.

The one escape hatch is WireGuard's optional PSK mode. Mixing a 256-bit pre-shared key into the Noise handshake makes the derived symmetric keys independent of Curve25519: even if the EC discrete log is solved, the PSK contribution keeps the keys secret. This would make $\seffconf{L_3} = \pqsafe$, blocking exposure at depth~2, reducing $d^*$ to~1, and, perhaps more importantly, restoring the VPN's privacy function against a quantum adversary. However, PSK mode requires out-of-band key distribution and standard deployments do not enable it.

%Table~\ref{tab:cross} summarizes the chain-level results across all four scenarios.

% \begin{table*}[t]
% \centering
% \caption{Cross-scenario comparison of chain-level PQC status and HNDL exposure.}
% \label{tab:cross}
% \footnotesize
% \begin{tabular}{@{}lcccccl@{}}
% \toprule
% \textbf{Scenario} & $n$ & $\schainconf$ & $\schainauth$ & $\schainmeta$ & $d^*$ & \textbf{Payload HNDL} \\
% \midrule
% 1: iMessage (WPA3+TLS+PQ3) & 3 & \pqsafe & \pqunsafe & \pqunsafe & 2 & No \\
% 2: HTTPS (WPA2-PSK+TLS) & 2 & \pqunsafe & \pqunsafe & \pqunsafe$^\dagger$ & 2 & Yes \\
% 3: HTTPS (WPA2-Ent+TLS) & 2 & \pqunsafe & \pqunsafe & \pqunsafe & 2 & Yes \\
% 4: HTTPS (WPA3+WG+TLS) & 3 & \pqunsafe & \pqunsafe & \pqunsafe & 3 & Yes \\
% \bottomrule
% \end{tabular}
% \end{table*}

\section{Discussion}
\label{discussion}
 We summarize the findings that emerge from these case studies. 

\noindent\textbf{Finding 1: One PQ-safe layer suffices for payload confidentiality.} Scenario~1 achieves $\schainconf = \pqsafe$ with a single PQ-safe layer (PQ3 at L7) despite two \pqunsafe{} layers below it. Scenarios~2--4, with zero PQ-safe layers, all have $\schainconf = \pqunsafe$. This validates the confidentiality sufficiency principle (RC2).

\noindent\textbf{Finding 2: Authentication is universally \pqunsafe.} Across all four scenarios and all nine active layer instances, no layer uses post-quantum digital signatures. Even PQ3 signs with ECDSA-P256. This confirms the authentication necessity concern (RC3).

\noindent\textbf{Finding 3: More layers $\neq$ better quantum security.} Scenario~4 ($n = 3$) achieves the worst HNDL outcome ($d^* = 3 = n$). Scenario~1 ($n = 3$) achieves $d^* = 2$ because one layer is \pqsafe. Layer count is irrelevant; what matters is whether any layer is \pqsafe.

\noindent\textbf{Finding 4: Metadata exposure halts at the first PQ-safe layer.} In Scenario~1, the adversary penetrates depths~1--2 but is blocked at depth~3 (PQ3). In Scenario~4, no layer blocks, and the adversary reaches application data. The exposure depth $d^*$ is determined entirely by the position of the first \pqsafe{} layer counting inward from the outermost.

\noindent\textbf{Finding 5: Enterprise security inversely correlates with quantum security.} Comparing Scenarios~2 and~3 (Table~\ref{tab:cs2v3}): WPA2-PSK is \pqunsafe$^\dagger$ (Grover, config fix) while WPA2-Enterprise is \pqunsafe{} (Shor, protocol replacement). The classically stronger protocol is quantum-weaker with a harder remediation path.

\noindent\textbf{Finding 6: Trust boundaries mirror quantum exposure boundaries.} The endpoint tables (Tables~\ref{tab:cs1-endpoint}--\ref{tab:cs4-endpoint}) reveal a pattern: at each intermediate node, the HNDL exposure column often recovers the same data that the classical exposure column already shows. In Scenario~1, Apple's relay classically sees iMessage metadata; an HNDL adversary recovers the same metadata and nothing more. In Scenario~4, the VPN server classically sees browsing destinations; an HNDL adversary at the VPN server can additionally break TLS to recover content (which the VPN server cannot do classically). The endpoint analysis thus reveals both cases where HNDL \emph{extends} classical exposure to content (CS2, CS3, CS4) and cases where a \pqsafe{} layer ensures HNDL provides \emph{no advantage beyond} what intermediaries already see (CS1).

\noindent\textbf{Finding 7: The $\dagger$ distinguishes fixable from structural vulnerabilities.} Both Scenarios~2 and~3 have $d^* = 2$ and $\schainconf = \pqunsafe$, but their L2 vulnerability mechanisms differ. WPA2-PSK's Grover-based vulnerability is eliminated by upgrading to AES-256 (a configuration change). WPA2-Enterprise's Shor-based vulnerability and WPA3-SAE's Shor-based vulnerability both require protocol redesign. The formal model assigns the same status (\pqunsafe) to both, but the remediation cost differs by an order of magnitude.

\section{Conclusion}
\label{sec:conclusion}
Existing work~\cite{baseri2024quantumsafeprotocols} on post-quantum migration has largely focused on standards guidance and protocol-specific redesigns, such as hybrid TLS key exchange, post-quantum secure messaging handshakes, and post-quantum VPN variants. In contrast, we study how the PQC posture of multiple concurrently active layers composes across a single communication stack, and how that composition differs by security objective such as confidentiality, authentication, and metadata exposure. We have presented a preliminary framework to analyze PQC threat across the network stack and demonstrated how to use this framework using some case studies.

% REFERENCES
\bibliographystyle{IEEEtran}
\bibliography{references}

\end{document}